\documentclass[landscape,usenatbib,usegraphicx]{mn2e}
\newcommand{\be}{\begin{eqnarray}}
\newcommand{\ee}{\end{eqnarray}}

\def\theequation{\thesection.\arabic{equation}}
\title[CMB constraints on radiative feedback]
{CMB polarization constraints on radiative feedback}

\author[C.~Burigana et al.]
{\parbox[t]{\textwidth}{
C.~Burigana \thanks{burigana@iasfbo.inaf.it},$^1$ 
L.A.~Popa \thanks{lpopa@venus.nipne.ro},$^{2,1}$
R.~Salvaterra \thanks{salvaterra@mib.infn.it},$^3$
\\
R.~Schneider \thanks{raffa@arcetri.astro.it},$^{4}$ 
T.~Roy~Choudhury \thanks{chou@ast.cam.ac.uk},$^5$ and
A.~Ferrara \thanks{ferrara@sissa.it}$^6$}
\\
\vspace*{3pt} \\
$^1$ INAF-IASF Bologna, Istituto di Astrofisica Spaziale e Fisica Cosmica di Bologna, \\
Istituto Nazionale di Astrofisica, via Gobetti 101, I-40129 Bologna - Italy \\
$^2$ Institute for Space Sciences, Bucharest-Magurele, Str. Atomostilor, 409, 
PoBox Mg-23, Ro-077125, Romania \\
$^3$Dipartimento di Fisica G.~Occhialini, Universit\`a degli Studi di Milano Bicocca, 
Piazza della Scienza 3, 
I-20126 Milano, Italy \\
$^4$ INAF - Osservatorio Astrofisico di Arcetri, Largo Enrico Fermi 5, 
I-50125 Firenze, Italy \\
$^5$ Institute of Astronomy, Madingley Road, Cambridge CB3 0HA, UK\\
$^6$ SISSA/International School for Advanced Studies, Via Beirut 4, 
I-34100 Trieste, Italy
}

\begin{document}

\def\lsim{\,\lower2truept\hbox{${< \atop\hbox{\raise4truept\hbox{$\sim$}}}$}\,}
\def\gsim{\,\lower2truept\hbox{${> \atop\hbox{\raise4truept\hbox{$\sim$}}}$}\,}

\maketitle

\begin{abstract}
We compute the imprints left on the CMB by two cosmic reionization models consistent 
with current observations but characterized by alternative radiative feedback prescriptions
({\it suppression} and {\it filtering}) resulting in a different suppression of star formation in 
low-mass halos. 
The models imply different ionization and thermal histories and 21 cm background signals. 
The derived Comptonization, $u$, and free-free distortion, $y_B$, parameters are 
below current observational limits for both models. However, the value of $u \simeq 1.69 \times 
10^{-7}$ ($\simeq 9.65 \times 10^{-8}$) for the suppression (filtering) model 
is in the detectability range of the next generation of CMB spectrum experiments.
Through the dedicated Boltzmann code CMBFAST, modified to include 
the above ionization histories, we compute the CMB angular power spectrum
(APS) of the TT, TE, and EE modes. For the EE mode the differences
between these models are significantly larger than the cosmic and sampling
variance over the multipole range $\ell \sim 5$~−-~15, leaving a good chance of discriminating 
between these feedback mechanisms with forthcoming/future
CMB polarization experiments. The main limitations come from
foreground contamination: it should be subtracted at per cent level in terms of
APS, a result potentially achievable by novel component separation techniques
and mapping of Galactic foreground.
\end{abstract}


\raggedbottom
\setcounter{page}{1}
\section{Introduction}
\setcounter{equation}{0}
\label{intro}

The accurate understanding of the ionization history of the universe
plays a fundamental role in the modern cosmology.
The classical theory of hydrogen
recombination for pure baryonic cosmological models
\citep{peebles, zeldovich} has been
subsequently extended to non-baryonic dark matter models
\citep{zabotin, jones, recfast} and
recently accurately updated to include also helium 
recombination in the current cosmological scenario 
(see \cite{switzer_hirata_07_3} and references therein).
Various models of the subsequent universe ionization history 
have been considered to take into account
additional sources of photon and energy production, possibly associated to
the early stages of structure and star formation, able to significantly
increase the free electron fraction, $x_e$,
above the residual fraction ($\sim10^{-3}$)
after the standard recombination epoch
at $z_{\rm rec} \simeq 10^3$.
These photon and energy production processes associated to this reionization 
phase may leave imprints in the cosmic microwave background (CMB) providing a crucial 
``integrated'' information 
on the so-called {\it dark} and {\it dawn} {\it ages}, i.e. the epochs
before or at the beginning the formation of first cosmological structures.
For this reason, among the extraordinary results achieved by 
the {\it Wilkinson Microwave Anisotropy Probe} (WMAP)
mission,
the contribution to the understanding of the cosmological 
reionization process has received a great attention.

To first approximation, the beginning of the reionization process is identified
by the Thomson optical depth,
$\tau$. The values of $\tau$ compatible with WMAP 3yr data, possibly complemented 
with external data, are typically in the range $\sim 0.06-0.12$ (corresponding to
a reionization redshift in the range $\sim 8.5-13.5$ for a sudden reionization 
history), the exact interval depending on the
considered cosmological model and combination of data sets \citep{Spergel2007}. 
While this simple ``$\tau$-parametrization'' of the reionization process and, in particular,
of its imprints on the CMB anisotropy likely represents a sufficiently accurate modelling
for the interpretation of current CMB data, a great attention has been recently posed on the 
accurate computation of the reionization signatures in the CMB for
a large variety of astrophysical scenarios and physical processes
(see e.g. \cite{psh,doroshkevich02,cen2003,ciardi03,doroshkevich03,kasuya,hansen}; 
Popa, Burigana \& Mandolesi (2005); \cite{wyithe})
also in the view of WMAP accumulating data and of forthcoming and future experiments beyond WMAP.
In this context, this work represents a step forward of our previous paper
\citep{paperI} dedicated to the study of the impact of reionization, and the associated 
radiative feedback, on galaxy formation and of the corresponding detectable signatures.
In that work we carried out a detailed comparison of two well defined alternative prescriptions
({\it suppression} and {\it filtering})
for the radiative feedback mechanism suppressing star formation in small-mass halos 
showing that 
they are consistent with a wide set of existing observational data but predict 
different 21 cm background signals accessible to future observations.
We focus here specifically on the signatures detectable in the CMB.

We observe that all viable (i.e. data-consistent)
reionization models require radiative feedback.
In fact, in the absence of radiative feedback
Population (Pop) III stars would be allowed to form efficiently down to low redshifts, 
and, in order to match the low-$z$ Gunn-Peterson data, a decreased Pop III 
star formation efficiency is required, in turn yielding a too low
Thomson optical depth\footnote{The filtering model implies a value of 
$\tau$ lower than that obtained in the suppression model
(as explained in the next Section); decreasing 
it further would push the value of $\tau$ below the 1$\sigma$ limit 
set by WMAP 3yr data.}.
In other words, some feedback mechanism is necessary
to have enough photons at high-$z$ and to
avoid at the same time an excess of photons at late times.
Our paper indicates how to discriminate between the
two most physical radiative feedback prescriptions with future CMB data.

The paper is organized as follows. In Section \ref{models} we briefly summarize the main 
theoretical and observational aspects of the considered models focussing on those 
relevant for the analysis of the features in the CMB. In Section \ref{cmb} we present the results of
our computation for the CMB spectral distortions (Section \ref{spectrum}) and anisotropies in total intensity and 
polarization (Section \ref{anisotropies}) and compare them with the foreground and sensitivity limitation
of future CMB anisotropy missions (Section \ref{fore_planck}). 
Finally, Section \ref{conclusion} summarizes our conclusions.

Through this paper we assume a flat $\Lambda$CDM cosmological model
consistent with WMAP described by
matter and cosmological constant (or dark energy) density parameters 
$\Omega_m=0.24$ and $\Omega_\Lambda=0.76$, 
reduced Hubble constant $h=H_0/(100{\rm km/s/Mpc)}=0.73$, 
baryon density $\Omega_b h^2=0.022$, density contrast
$\sigma_8=0.74$, and adiabatic scalar
perturbations (without running) 
with spectral index $n_s=0.95$. We assume a CMB background temperature of 
2.725~K \citep{mather99}.

\section{Reionization models}
\setcounter{equation}{0}
\label{models}
The latest analysis of Ly$\alpha$ absorption in the
spectra of the 19 highest redshift Sloan Digital Sky Survey (SDSS) quasars (QSOs)
shows a strong evolution of the Gunn-Peterson Ly$\alpha$ opacity at $z \sim 6$ \citep{Fan2006, Gallerani2006}.
The downward revision of the
electron scattering optical depth to $\tau = 0.09 \pm 0.03$ in the release of
the 3-yr WMAP data \citep{Page2007,Spergel2007},
is consistent with ``minimal reionization models'' which do not require the presence of
very massive ($M>100 M_\odot$) Pop III stars \citep{Choudhury2006,Gnedin2006}.
We can then use these models
to explore the effects of reionization on galaxy formation, to which
we will refer to as ``radiative feedback".

Recently, Choudhury \& Ferrara (2005, 2006) 
developed a semi-analytic model to jointly study cosmic
reionization and the thermal history of the intergalactic medium
(IGM). According to \citet{paperI}, the semi-analytical model developed
by \citet{Choudhury2005} complemented by the additional physics introduced in \citet{Choudhury2006} 
involves: $i)$ the treatment of IGM inhomogeneities by adopting
the procedure of Miralda-Escud{\'e}, Haehnelt \& Rees (2000); $ii)$ the 
modelling of the IGM treated as a multiphase medium, following the thermal
and ionization histories of neutral, HII, and HeIII regions simultaneously
in the presence of ionizing photon sources represented by Pop III stars with a standard Salpeter IMF
extending in the range $1-100~M_\odot$ \citep{Schneider2006}, Pop II stars with $Z=0.2 Z_{\odot}$
and Salpeter IMF, and QSOs (particularly relevant at $z \lsim 6$);
$iii)$ the chemical feedback controlling the prolonged transition from Pop III to Pop II stars
in the merger-tree model by \cite{Schneider2006};  
$iv)$ assumptions on the escape fractions of ionizing photons, considered to be independent of
the galaxy mass and redshift, but scaled to the amount of produced ionizing photons.
It then accounts for radiative feedback inhibiting star formation in low-mass galaxies. 
This semi-analytical model is determined by only four free parameters: 
the star formation efficiencies of Pop II
and Pop III stars, a parameter, $\eta_\mathrm{esc}$, related to the escape fraction of ionizing
photons emitted by Pop II and Pop III stars, and the normalization of the photon
mean free path, $\lambda_0$, set to reproduce low-redshift observations of Lyman-limit 
systems \citep{Choudhury2005}.

A variety of feedback mechanisms can suppress star formation in mini-halos, 
i.e. halos with virial temperatures $T_{vir} < 10^4$~K, particularly if their 
clustering is taken into account \citep{Kramer2006}. We therefore assume that stars 
can form in halos down to a virial temperature of $10^4$~K, consistent with the 
interpretation of the 3-yr WMAP data 
(Haiman \& Bryan 2006; but see also Alvarez et al. 2006). 
Even galaxies with virial temperature $T_{vir} \gsim 10^4$~K can be significantly 
affected by radiative feedback during the reionization process, as the increase
in temperature of the cosmic gas can dramatically suppress their formation.
Based on cosmological simulations
of reionization, \citet{Gnedin2000} developed an accurate characterization of the radiative 
feedback on low-mass
galaxy. This study shows that the effect of photoionization is
controlled by a single mass scale in both the linear and non-linear regimes.
The gas fraction within dark matter halos  at any given moment is fully specified
by the current filtering mass, which directly corresponds to the length
scale over which baryonic perturbations are smoothed in linear theory. The
results of this study provide a quantitative description of radiative feedback,
independently of whether this is physically associated to photoevaporative flows
or due to accretion suppression.

Following \citet{paperI}, we consider here two specific alternative prescriptions
for the radiative feedback by these halos:

\noindent
$i)$ {\it suppression model} -- following \citet{Choudhury2006}, 
we assume that in photoionized regions halos  can form stars
{\it only} if their circular velocity exceeds the critical value
$v_\mathrm{crit} = {2 k_B T}/{\mu m_\mathrm{p}}$; 
here $\mu$ is the mean molecular weight, $m_\mathrm{p}$ is the proton mass, 
and $T$ is the average temperature of
ionized regions, computed self-consistently from the multiphase IGM model;

\noindent
$ii)$ {\it filtering model} -- following \citet{Gnedin2000}, we assume that 
the average 
baryonic mass $M_{b}$ within halos  
in photoionized regions
is a fraction of the universal value $f_{b} = \Omega_b/\Omega_m$, given by the fitting formula
${M_b}/{M} = {f_b}/{[1+(2^{1/3}-1) M_\mathrm{C}/M]^3}$,
where $M$ is the total halo mass and $M_\mathrm{C}$ is the total mass of halos  
that on average retain 50\% of their gas mass.
A good approximation for the characteristic mass 
$M_\mathrm{C}$ is given by the linear-theory filtering mass,
$M_\mathrm{F}^{2/3} = ({3}/{a}) \int_0^a da^\prime M_\mathrm{J}^{2/3}(a^\prime)
\left[1-\left({a^\prime}/{a}\right)^{1/2}\right]$,
where $a$ is the cosmic scale factor,
$M_\mathrm{J} \equiv ({4 \pi}/{3}) \bar{\rho} \left({\pi c_\mathrm{s}^2}/{G \bar{\rho}}\right)^{3/2}$,
is the Jeans mass, $\bar{\rho}$ is the average total mass density of the Universe, 
and $c_\mathrm{s}$ is the gas sound speed.
%


The model free parameters are constrained by a wide range of observational
data.  \citet{paperI} reported the best fit choice of the above four parameters for these two 
models that well accomplish a wide set of astronomical observations, such as the 
redshift evolution of Lyman-limit absorption systems, the Gunn-Peterson and electron 
scattering optical depths, the cosmic star formation history, and number counts of 
high redshift sources in the NICMOS Hubble Ultre Deep Field.

The two feedback
prescriptions have a noticeable impact on the overall reionization history and the relative contribution
of different ionizing sources. In fact, although the two models predict similar global star formation
histories dominated by Pop II stars, the Pop III star formation rates have
markedly different redshift evolution. Chemical feedback forces Pop III stars to live preferentially
in the smallest, quasi-unpolluted halos  
(virial temperature $\gsim 10^4$~K, \citealt{Schneider2006}), which are
those most affected by radiative feedback.
In the suppression model, where star formation is totally suppressed below 
$v_\mathrm{crit}$, Pop III
stars disappear at $z \sim 6$; conversely, in the filtering model, where halos  
suffer a gradual reduction of the available
gas mass, Pop III stars continue to form at $z \lsim 6$, though with a declining rate.
Since the star formation and photoionization rate at these redshifts are observationally well constrained,
the star formation efficiency and escape fraction of Pop III stars need 
to be lower in the filtering model in
order to match the data. 
Therefore reionization starts at $z \lsim 15$ in the filtering model and only 16\% 
of the volume is 
reionized at $z=10$
(while reionization starts at $z \sim 20$ in the suppression model and it is 85\% 
complete by $z=10$).
For $6 < z < 7$, QSOs, Pop II and Pop III give a comparable contribution 
to the total photo-ionization rate in the filtering model, whereas in the 
suppression model reionization at 
$z < 7$ is driven 
primarily by QSOs, with a smaller
contribution from Pop II stars only. 

\begin{figure}
\hskip -0.3cm
\includegraphics[width=6.3cm,angle=90]{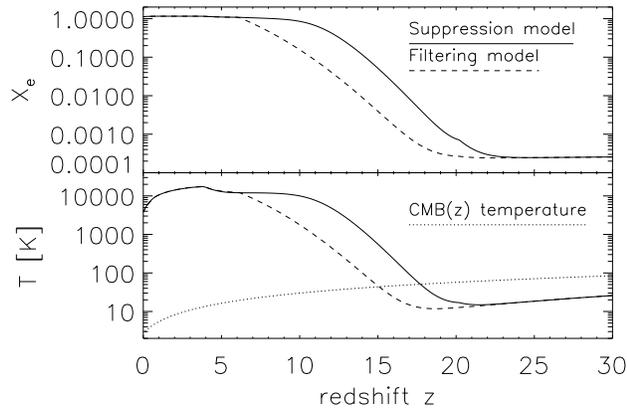}
\caption{Redshift dependence of the fraction of free electrons (top panel)
and of the gas kinetic temperature (bottom panel) for the two considered models.
See also the text.}
\label{fig:histories}
\end{figure}

The predicted free electron fraction and gas temperature evolution 
(see Fig.~\ref{fig:histories}) in the redshift
range $7 < z < 20$ is very different for the two feedback models.
Bottom panel of Fig.~\ref{fig:histories} compares the evolution of the gas kinetic temperature
and CMB temperature for the two models. 
In particular, in the filtering model the gas kinetic temperature is heated above 
the CMB value only at $z \lsim 15$. 

The Thomson optical depth, $\tau = \int \chi_e n_e \sigma_T c dt$,
can be directly computed for the assumed $\Lambda$CDM cosmological model
parameters given the ionization histories shown in the top panel of 
Fig.~\ref{fig:histories}. 
We find $\tau \simeq 0.1017$ and $\tau \simeq 0.0631$ 
for the suppression and the filtering
model, respectively.
Note that these values are consistent with the Thomson optical depth range 
derived from WMAP 3yr data \citep{Spergel2007}
but with $\sim 1 \sigma$ difference among the two models, leaving a chance of
probing them with forthcoming CMB anisotropy experiments.

\section{Signatures on the CMB}
\setcounter{equation}{0}
\label{cmb}

The cosmological reionization leaves imprints on the CMB depending on the
(coupled) ionization and thermal history. 
They can be divided in three categories\footnote{Inhomogeneous reionization 
also produces CMB secondary anisotropies that dominate over
the primary CMB component for $l\gsim 4000$ and can  be detected by upcoming
experiments, like the Atacama Cosmology Telescope or ALMA 
\citep{salvaterra,iliev}.}: 
$i)$ generation of CMB Comptonization and free-free spectral distortions
associated to the IGM electron temperature incresase during the reionization epoch,
$ii)$ suppression of CMB temperature anisotropes at large multipoles, $\ell$,
due to photon diffusion, and $iii)$ increasing
of the power of CMB polarization and temperature-polarization
cross-correlation anisotropy at various multipole ranges,
mainly depending on the reionization epoch,
because of the delay of the effective last scattering surface.

The imprints on CMB anisotropies are mainly dependent
on the ionization history while
CMB spectral distortions strongly depend
also on the thermal history.

It is interesting to compare the imprints left
in the CMB for the two models considered here to understand if 
they could be distinguished by forthcoming and future experiments.

\subsection{Spectral distortions}
\label{spectrum}

The Compton scattering of CMB photons with the electrons 
heated during the reionization process implies the generation
of a global Comptonization distortion \citep{ZS69, ZIS72} 
characterized by the Comptonization parameter 
$u \simeq (1/4) \Delta \epsilon / \epsilon_i$,
where $\Delta \epsilon / \epsilon_i$ is the fractional amount of 
energy exchanged between matter and radiation. 
In addition, we expect also the generation of a free-free distortion,
physically coupled to the Comptonization one, because of 
the bremsstrahlung photon production process 
\citep{Bref1, Bref2} in the hot gas. Its amplitude is characterized
by the so-called free-free distortion parameter, $y_B$.
For late processes, as in this case, 
the resulting distorted spectrum can be fully described
to a very good precision with the analytical formalism described in 
\cite{BDD95} based on the computation of the Comptonization
and free-free distortion parameters, by simply modifying \citep{buriganaetal04a}
the cosmic expansion time to take into account the cosmological constant
(or dark energy) contribution, dominant at low redshifts.
By exploiting the ionization and thermal histories shown in 
Fig.~\ref{fig:histories} we find
$u \simeq 1.69\times10^{-7}$, $y_B \simeq 9.01\times10^{-10}$ and 
$u \simeq 9.65\times10^{-8}$, $y_B \simeq 5.24\times10^{-10}$ respectively
for the suppression and the filtering model: these values are clearly well
below the COBE/FIRAS limits \citep{fixsen96,salvaterraburigana}.
The two models show similar ionization and thermal histories at $z \lsim 6$
while important differences are predicted at $z \gsim 6$. These explain the
different spectral distortion levels generated in the two cases.
In fact, considering only redshifts $z \gsim 6$, we find
$u \simeq 7.98\times10^{-8}$, $y_B \simeq 4.37\times10^{-10}$ and 
$u \simeq 1.05\times10^{-8}$, $y_B \simeq 8.26\times10^{-11}$ respectively
for the suppression and the filtering model.
We note that, while the free-free distortion levels predicted in these models
are too small to be detected even by long wavelength 
(i.e. at $\lambda \gsim 1$~cm)
spectrum experiments with precision comparable to -- or even significantly better 
than -- that of COBE/FIRAS \citep{KOG96,BS03}, 
such Comptonization distortion prediction levels are comparable to 
those that could be in principle observed by a future generation of 
CMB spectrum experiments, in particular at millimetre wavelengths, able to improve by 
$\approx 30-100$ times the COBE/FIRAS results \citep{FM02,buriganaetal04b}.
In this perspective, future accurate measurements at long wavelengths 
\citep{luna}
could significantly improve the reliability of the detection of such Comptonization 
distortions, removing the approximate degeneracy in the joint
determination of early (Bose-Einstein like, \cite{BE}) and late distortion parameters 
that remains in the presence of accurate spectrum measures only at millimetre 
wavelengths.

\subsection{Anisotropies}
\label{anisotropies}

\begin{figure}
\hskip -12.cm
\vskip 0.15cm
\includegraphics[width=10.cm,height=9.cm,angle=90]{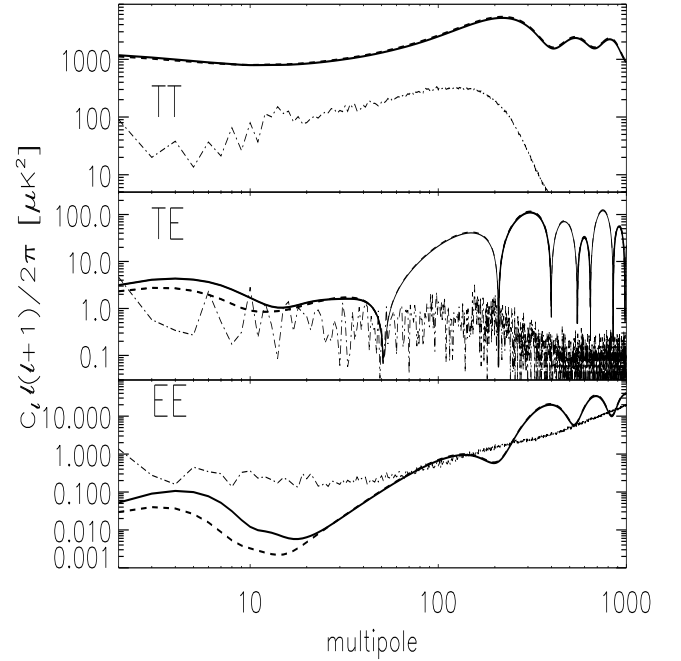}
\vskip -0.8cm
\caption{APS of CMB anisotropies
for the three considered modes TT, TE, EE, reported in each panel and
computed by introducing 
the ionization and thermal histories shown in Fig.~\ref{fig:histories}
in the {\tt CMBFAST} code
(solid lines: suppression model; dashes: filtering model). 
Thick lines denote correlation, while thin lines denote anticorrelation 
(appearing for the TE mode at $\ell \gsim 50$).
The APS of the foreground, dominated by the Galactic contribution, 
is reported for comparison (dot-dashes). 
The WMAP 3yr V band data are considered here.
See also the text. [Results expressed in terms of thermodynamic temperature fluctuations].}
\label{fig:aps}
\end{figure}

The detailed computation of the reionization imprint 
in the CMB anisotropy angular power spectrum (APS), $C_\ell$, 
requires the use of dedicated Boltzmann 
codes, properly implemented to have the possibility
of introducing the adopted ionization history.
We have modified the public version (4.5.1) of 
the {\tt CMBFAST} code\footnote{http://www.cmbfast.org/}
(see e.g. \cite{seliak_zalda_96})
to properly replace \citep{popa05}
the simple step function (or Heaviside function) approximation for the ionization fraction
adopted there to model the reionization process
with the considered ionization histories, shown in Fig.~\ref{fig:histories}.
Our results 
are reported in Fig.~\ref{fig:aps}. Having neglected for simplicity
tensor perturbations, we focus here on the TT (total intensity, i.e. temperature), 
TE (temperature-polarization cross-correlation), and EE polarization modes of
the CMB anisotropy APS. 
In Fig.~\ref{fig:aps} we display also the APS of the foreground 
in the V band (centred at 61 GHz) of WMAP 3yr 
data\footnote{http://lambda.gsfc.nasa.gov/product/map/current/},
a frequency channel where the foreground is found to be minimum
(or almost minimum) in both temperature and polarization
\citep{bennettfore,Page2007} at the angular scales larger than $\sim 1^\circ$
of relevance in this context. 
It is simply derived with the {\tt anafast} facility of the {\tt HEALPix} 
\footnote{http://healpix.jpl.nasa.gov/}
package
\citep{gorski05}
analysing the difference between the original map and the CMB map
and adopting the mask used by the WMAP team for the polarization 
analysis (covering $\simeq 74$ per cent of the sky).
For sake of simplicity, 
no attempt is made in these foreground plots
to subtract the noise contribution 
to the overall APS\footnote{The relevance of the noise term 
-- important at high multipoles -- obvioulsy 
increases going from the TT, to the TE, and then to the EE mode
because of the decreasing of the signal-to-noise ratio.}.
We are in fact interested here to low multipoles
($\ell \lsim$~few tens)
where the differences between these two models show up, 
as expected for relatively late reionization processes, and the signal power 
dominates over the noise one.
Note that the relative difference between the APS
computed for the suppression and filtering model
as well as the foreground impact compared to the CMB signal 
increase going from the TT, to the TE, and then to the EE mode.

\subsection{Foreground and perspectives from Planck and future polarization experiments}
\label{fore_planck}

To understand if these classes of models can be distinguished with forthcoming and future CMB 
anisotropy experiments we compare their relative difference with 
the cosmic and sampling 
variance limitation, $\Delta C_\ell/C_\ell$, \citep{knox95}, i.e. neglecting the
contribution by the instrumental noise, 
and a potential residual foreground contamination.
At the low multipoles ($\ell \lsim 20-30$) relevant here (see
Fig.~\ref{fig:aps})
the {\it direct} (ideal) instrumental noise limitation
is in fact not critical for current and future space missions.
On the contrary,
the experiment sensitivity and frequency coverage and, in particular, the rejection (by 
instrument design and/or by subsequent data analysis) of the various classes of systematic 
effects are critical to have the possibility 
to accurately subtract the foreground contribution 
and to effectively achieve the ultimate cosmic variance limit  
(see e.g. \cite{mennella} and \cite{popaNAR}, and references therein, 
for reviews devoted to these aspects in the context of the {\it 
Planck} mission).
So, the results reported here apply to the case of the forthcoming {\it Planck} 
mission\footnote{http://www.rssd.esa.int/planck}, at least in the best case of 
an effective excellent rejection of all the potential systematic effects, and to the
next generation of CMB polarization anisotropy experiments with a (nearly) all-sky coverage
studied in the NASA (Inflation Probe, CMBPol)
and the ESA 
(B-Pol\footnote{http://www.th.u-psud.fr/bpol/index.php})
context.

\begin{figure}
\hskip -0.7cm
\includegraphics[width=8cm,height=9.cm,angle=90]{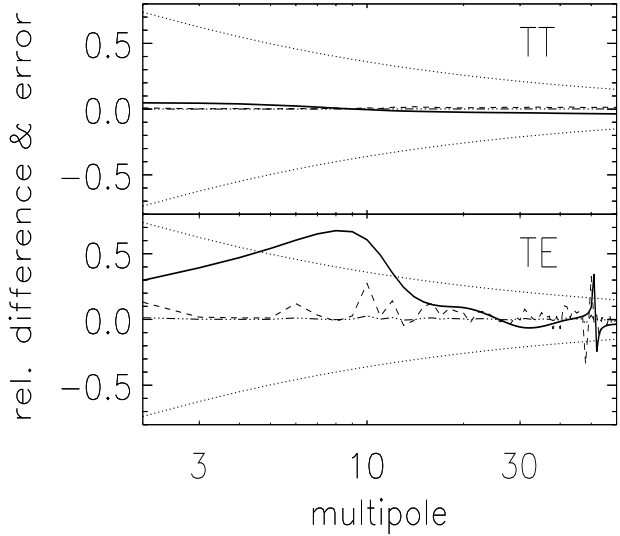}
\vskip -0.3cm
\caption{Relative difference, $\delta^{S,F}$, 
between the (TT and TE mode) APS of CMB anisotropies 
for the suppression and filtering models
reported in Fig.~\ref{fig:aps} (thick solid lines)
compared with the cosmic and sampling variance limitation
corresponding to a sky coverage of $\simeq 74$ per cent
(region between the dotted lines).
We report for comparison the APS from a potential residual foreground
(dot-dashes)
corresponding to two different values of $f$: 0.1 (dashed line) 
and 0.01 (three dots-dashed line).
See also the text.}
\label{fig:diff_TT_TE}
\end{figure}

\begin{figure}
\hskip -0.5cm
\includegraphics[width=8cm,height=9.cm,angle=90]{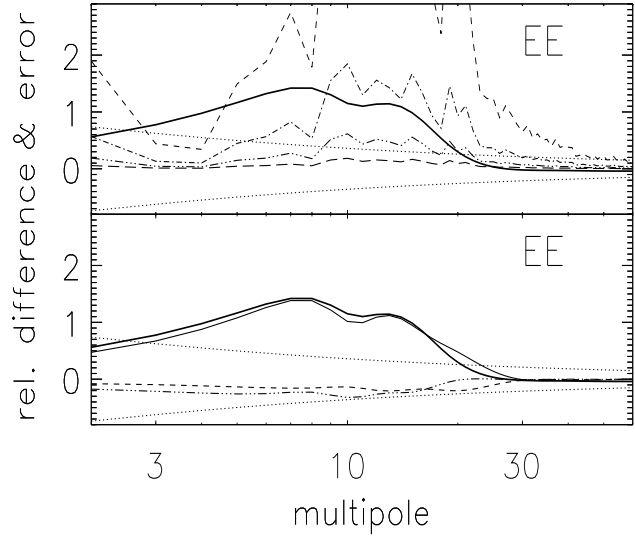}
\vskip -0.3cm
\caption{Relative difference, $\delta^{S,F}$, 
between the (EE mode) APS of CMB anisotropies for the suppression and filtering models
reported in Fig.~\ref{fig:aps} (thick solid lines)
compared with the cosmic and sampling variance limitation
corresponding to a sky coverage of $\simeq 74$ per cent
(region between the dotted lines).
In the top panel
we report also for comparison the APS from a potential residual foreground
(dot-dashes)
corresponding to different values of $f$: 0.1 (dashed line), 0.03 (dots-dashed line), 
0.01 (three dots-dashed line), and 0.003 (long dashes).
In the bottom panel
we show also $\delta^{SH,FH}$ computed by replacing
the two considered models with those obtained assuming the 
simple step function
implemented in the public {\tt CMBFAST} code for the reionization history
(thin solid lines)
for the corresponding values of optical depth.
The relative difference between the results based on the suppression (resp. 
filtering) model and its step function approximation with
$\tau \simeq 0.1017$ (resp. with $\tau \simeq 0.0631$) 
is shown by the dashed line (resp. three dots-dashed line). 
See also the text.}
\label{fig:diff_EE}
\end{figure}

Following \cite{nc04}, we compute the quantity 
$\delta^{M,N} = (C_\ell^M-C_\ell^N)/[0.5(C_\ell^M+C_\ell^N)]$
where the index $M$ or $N$ specifies the APS calculated for a given model (for simplicity, we 
omit here the indices TT, TE or TE; they are reported in each figure panel).
A potential foreground residual contamination,
resulting from a non-perfect (possibly biased) component separation, is modelled here as 
$\delta^{Fore} = f C_\ell^{Fore}/[0.5(C_\ell^M+C_\ell^N)]$, where $C_\ell^{Fore}$
is the foreground APS and the {\it effective} parameter $f$ characterizes  
the accuracy of the component separation method in the considered range of multipoles.

Figs.~\ref{fig:diff_TT_TE} and \ref{fig:diff_EE} summarize our results. We display 
$\delta^{M,N}$ for
$M=S$ (suppression model) and $N=F$ (filtering model). In the case of the EE mode, we 
report also the results obtained 
exploiting the Heaviside function approximation of the two reionization histories
with corresponding optical depths, $M=SH$ and $N=FH$. It is also interesting
to consider $\delta^{M,N}$ for $M=S$ and $N=SH$ and for $M=F$ and $N=FH$ in order to
quantify the relative error implied by the step function approximation
of a more complex time-dependent reionization history.

As evident from Fig.~\ref{fig:diff_TT_TE} (top panel), the difference between the two models 
is overwhelmed by the cosmic and sampling variance
in the case of temperature anisotropy measures.
The difference in the TE mode (see bottom panel of Fig.~\ref{fig:diff_TT_TE}) 
turns to be well above the foreground limitation 
even for a removal accuracy of about 10 per cent. Unfortunately, it is above the cosmic 
and sampling variance
limitation by a factor of $\sim 1.5$ only and for a small range of multipoles 
($\ell \sim 7-10$)\footnote{The apparently large features at $\ell \sim 50$
result from the very small absolute value of the TE mode close to 
its change of sign with the consequence of large relative differences 
between the two models but only on a very small multipole range 
(see also Fig.~\ref{fig:aps}).}. 
On the contrary, for the polarization EE mode APS 
the difference between the two considered models is significantly larger 
than the cosmic and sampling variance over a interesting range of multipoles
($\ell \sim 5-15$), as shown by Fig.~\ref{fig:diff_EE} (top panel). 
In this case the main limitation derives from a possible
residual foreground contamination: as evident, a foreground removal 
accuracy at per cent level (or better)
in terms of APS is necessary to accurately exploit the 
information contained in CMB polarization about the cosmological reionization process.
Note that, possibly except for some sky areas at very low foreground contamination level
\citep{carrettietal06},
similar (resp. or better) accuracies are needed for the observation of BB polarization 
mode of the CMB anisotropy\footnote{http://www.b-pol.org/pdf/BPOL\_Proposal.pdf}
for ratios of tensor to scalar perturbations 
$\lsim {\rm some} \times 10^{-3}$ 
on typical sky regions at low/moderate foreground contamination level  
(resp. at low multipoles) \citep{laportaetal06}.
This calls for a further progress in component separation techniques in polarization
(see e.g. \cite{baccigalupietal04,stivoli,Page2007,aumont} for promising developments)
and for an accurate mapping of the (mainly Galactic) polarized foregrounds 
in radio
and far-IR bands\footnote{See e.g. the talks by J.P.~Leahy and
J.-P.~Bernard at\\
http://www.th.u-psud.fr/bpol/talks\_orsay\_meeting/Talks.html} 
to complement microwave surveys by improving
both the foreground physical modelling 
and the component separation accuracy.

Finally, we observe that, particularly at $\ell \simeq 10-20$, the step function approximation
of the reionization history adopted in the simple $\tau$-parametrization,
although adequate to reveal the bulk of the difference between these models,
implies an error on the EE mode comparable to the cosmic and sampling variance\footnote{A 
similar analysis for the TT (resp. TE) mode shows that this kind of error is overwhelmed 
by (resp. about one order of magnitude less than)
the cosmic and sampling variance.}
(see bottom panel of Fig.~\ref{fig:diff_EE}).
Therefore,
a proper treatment of the cosmological reionization history in the Boltzmann
codes is necessary for a fine comparison with 
the high accuracy data from forthcoming and future CMB polarization anisotropy experiments.

\section{Conclusion}
\setcounter{equation}{0}
\label{conclusion}

We have analyzed the implications for the CMB of the self-consistent semi-analytical model 
developed by Choudhury \& Ferrara (2005, 2006)
to describe the effect of reionization
and its associated radiative feedback on galaxy formation, modelling the 
suppression of star formation in
low-mass galaxies due to the increase in temperature of the cosmic gas in ionized regions 
according to two different feedback prescriptions, the {\it suppression} and 
{\it filtering} model.
These two models predict different ionization and thermal histories, in particular at $z \gsim 6$,
resulting into Thomson optical depths
$\tau \simeq 0.1017$ and $\tau \simeq 0.0631$, respectively, 
consistent with the WMAP 3yr data.

According to the usual semi-analytical approach to model CMB spectral distortions,
we computed the Comptonization and free-free distortion parameters, $u$ and $y_B$.
Their are found to be well below the current observational limits.
In particular, the values obtained for the Comptonization parameter 
($u \simeq 1.69\times10^{-7}$ and $\simeq 9.65\times10^{-8}$
for the suppression and the filtering model, respectively)
are in range accessible to a future generation of
CMB spectrum experiments, in particular at millimetre wavelengths,
able to improve by $\approx 30-100$ times the COBE/FIRAS results.

We have modified the Boltzmann 
code {\tt CMBFAST} to introduce the considered ionization histories and 
compute the CMB angular power spectrum
(namely, the APS of temperature (TT), temperature-polarization correlation (TE), 
and polarization EE modes) 
improving the step function approximation 
adopted in the simple ``$\tau$-parametrization''.
The differences between the predictions of the two models are 
negligible for the TT mode and small for the TE mode when compared to 
cosmic and sampling variance limitation
but are of particular interest for the EE polarization mode.
They are significantly larger 
than the cosmic and sampling variance over the multipole range
$\ell \sim 5-15$, leaving a good chance of discriminating 
between the two feedback mechanisms
with forthcoming and future CMB polarization anisotropy experiments, 
and in particular in the view of the forthcoming ESA {\it Planck} satellite 
that will launched in about one year.
The main limitation derives from foreground contamination: it should be 
subtracted at per cent level in terms of APS,
a result achievable with joint efforts in component separation 
techniques and in the mapping and modelling of the Galactic foreground.
Also, a particular care should be taken in the control of potential 
systematic effects down to negligible levels
and in the precise reconstruction of the APS from the data at low multipoles.

Our analysis indicates that forthcoming CMB anisotropy polarization data 
together with future 21 cm data \citep{paperI}
will play a crucial role in breaking current
degeneracies and constraining the cosmological reionization process and 
the astrophysical properties of the sources responsible for it 
at high redshifts ($z \gsim 6$).

\bigskip
\bigskip
\noindent
{\bf Acknowledgements.} 
The use of the {\tt CMBFAST} code is acknowledged.
Some of the results in this paper have been derived using {\tt HEALPix}
\citep{gorski05}. 
The availability of the WMAP 3-yr maps is acknowledged.
We are grateful to Benedetta Ciardi for profitable discussions and to DAVID
members\footnote{http://www.arcetri.astro.it/science/cosmology} for fruitful
comments. We warmly thank the members of {\it Planck} Working Groups and Core Teams
for useful discussions. 
CB acknowledge the support by
the ASI contract ``Planck LFI Activity of Phase E2''.

\end{document}